\documentclass[preprint,showpacs,preprintnumbers,amsmath,amssymb]{revtex4}%
\usepackage{graphicx}
\usepackage{dcolumn}
\usepackage{bm}
\usepackage{amsmath}
\usepackage{amsfonts}
\usepackage{amssymb}%
\begin{document}

\title{The  generalized second law of thermodynamics for the interacting in $f(T)$ gravity}
\author{Ram\'on Herrera}
\email{ramon.herrera@ucv.cl}
\affiliation{Instituto de F\'{\i}sica, Pontificia Universidad Cat\'{o}lica de Valpara%
\'{\i}so, Casilla 4059, Valpara\'{\i}so, Chile. }

\begin{abstract}
We study the validity of the generalized second law (GSL) of
gravitational thermodynamics in a non-flat FRW universe containing
the interacting in $f(T)$ gravity. We consider  that the boundary
of the universe to be confined  by the dynamical apparent horizon
in FRW universe. In general, we discuss the effective equation of
state, deceleration parameter and GLS in this framewok. Also, we
find that the interacting-term $Q$ modifies these quantities and
in particular,  the evolution of the total entropy, results in an
increases on the GLS of thermodynamic, by a factor
$4\pi\,R_A^3\,Q/3$. By using a viable $f(T)$ gravity with an
exponential dependence on the torsion, we develop a model where
the  interaction term  is related to the total energy density of
matter. Here, we find that a crossing of phantom divide line is
possible for the interacting-$f(T)$ model.
\end{abstract}
\maketitle

\section{Introduction}

Observational data of the luminosity-redshift of type Ia
supernovae (SNeIa), large scale structure (LSS) and the cosmic
microwave background (CMB) anisotropy spectrum, have provided
confirmation that our universe has recently entered a phase of
accelerated expansion (\cite{ac}). A possible responsible of this
acceleration of the universe is the dark energy (DE)  and the
nature of this energy  is an important problem today in the modern
physics. For a review of DE candidates and models, see
Ref.(\cite{MDE}).

In the last years, a $f(T)$ theory was introduced to explain the
current expansion of the universe without considering the DE
(\cite{ft,ft1,ft2}). The $f(T)$ theory is a generalization of the
teleparallel gravity (TG) and becomes equivalent of General
Relativity (\cite{ft3}). The idea original of the $f(T)$ theory
results from the generalization of the TG from  the torsion scalar
$T$ (\cite{ft, ft4}), similarly  to the motivation of the Ricci
scalar $R$ in the Einstein-Hilbert action by replacing the
function $f(R)$. However, in $f(T)$ theory the field equations are
second order as opposed to the fourth order equations of $f(R)$
theory. In the formalism of the TG, the tetrad fields are the
basic variable of TG together with the Weitzenbock connection, see
Refs.(\cite{ft,ft3,ft5}) for a review.


In the recent years, the thermodynamics aspects of the
accelerating universe has considered much attention and different
results has been obtained (\cite{GLS1}). In particular, the
verification of the first and second law of the thermodynamic,
studying the dynamic together with the thermodynamics aspect of
the accelerated expansion of the universe.

In the context of the validity  the  generalized second law (GSL)
of thermodynamics, is necessary that the evolution with respect to
the cosmic time of the total entropy
$\dot{S}_{Total}=d(S_A+S_m)/dt\geq 0$. Here, $S_A$ is the
Bekenstein-Hawking entropy on the apparent horizon and $S_m$
denotes the entropy of the universe filled matter (pressureless
baryonic matter (BM) and dark matter (DM)) inside the dynamical
apparent horizon (\cite{hor}). Therefore, in conformity with the
GLS of thermodynamic, the evolution of the total entropy,
$S_{Total}$, cannot decrease in time (\cite{S1, S2}).

On the other hand, considering the first law of thermodynamics, we
can write $-dE=T_A\,dS_A$ to the apparent horizon  and obtain in
this form the Einstein's field equation. This equivalency is
satisfied, if we consider that the Hawking temperature $T_A\propto
R_A^{-1}$ and the entropy on the apparent horizon
 $S_A \propto A$, where  $R_A$ and $A$ are the radius and
area associated to the horizon (\cite{hor}), see also
Ref.(\cite{other}). However, the entropy on the apparent horizon
$S_A$, is modified for other types of theories. In particular, in
the case of $f(T)$ gravity, Miao \textit{et al.} in
Ref.(\cite{Miao}), calculated that when $f''$ is small, the
entropy of the apparent horizon resulted to be $S_A=A\,f'/4G$.
Also, for $f(R)$ gravity the entropy is changed, see
Ref.(\cite{fR}). Here, the primes denote derivative with respect
to the torsion scalar $T$.

In relation to the GLS of thermodynamics in the framework of
$f(T)$ gravity, was developed in Ref.(\cite{KAR}). Here, the
authors studied for a spatially flat universe, the validity of the
GLS for two viable $f(T)$ models; $f(T)=T+\mu_1(-T)^{n}$ and
$f(T)=T-\mu_2\,T\,(1-e^{\beta T_0/T})$, originally proposed in
Refs.(\cite{9,11}). The GLS in $f(T)$ gravity with entropy
corrections was studied in (\cite{uu}), where two different cosmic
horizon are analyzed. Also, the GLS in the emergent universe for
some viable models of $f(T)$ gravity was considered  in
(\cite{Ghosh}), were considered a non-flat universe for three
viable models of $f(T)$ gravity to an emergent universe (see also
Ref.(\cite{em})).

In order to solve the cosmic coincidence problem (\cite{Prob}),
various authors  have studied the interaction between DE and DM
components (\cite{Int}). The interaction term  can alleviate the
coincidence problem in the sense that the rate between the
densities of the DM and DE either leads to a constant or varies
slowly in late times (\cite{Int2}). In the context of the GLS, the
study of the validity of the GLS for the interaction between DM
and DE was analyzed in Ref. (\cite{Karami}). Here, the interaction
between both component is proportional to the DE. Also, the
thermodynamics description for the interaction between holography
and DM was studied in Ref. (\cite{Wang}) and an analysis of the
GLS for the interacting generalized Chaplygin gas model was
considered in Ref. (\cite{Ka1}). In Ref. (\cite{Pavon}) was
considered the interaction between DE and DM from the Le
Ch\^{a}telier-Braun principle. Also, a phase space analysis of
interacting dark energy in $f(T)$ cosmology was analyzed in Ref.
(\cite{cc}) and  the behavior of interacting Ricci dark energy in
logarithmic $f(T)$ gravity was studied in Ref. (\cite{cc2}).

In the present work we are interested in investigating the
validity of the GSL of thermodynamics in the context of the
interaction-$f(T)$ gravity. We considered a non-flat universe
Friedmann-Robertson-Walker (FRW) background occupied with the
pressureless matter. Also, we analyzed   the behaviors of the
equation of state (EoS),  together with the deceleration
parameter. As an example, we analyzed for a flat FRW universe, one
viable $f(T)$ model together with a particular interaction term.

The outline of the paper is as follows. The next section presents
the interacting $f(T)$ gravity. Here we study the EoS and the
deceleration parameter. Section \ref{g} we investigate the
validity of the GSL of thermodynamics. Section \ref{e} we develop
an example for $f(T)$ and the interaction term in a flat FRW
universe. Finally, in Sect.\ref{conclu} we summarize our finding.
We chose units so that $c=\hbar =1$.

\section{ Interacting $f(T)$ gravity}\label{f}

The action $I$ of modified TG in the framework of $f(T)$ gravity,
becomes\cite{ft}
\begin{equation}
I=\frac{1}{16\pi\, G}\int\,d^4x\,\sqrt{-g}\,[f(T)+L_m].
\end{equation}
Here $L_m$ is related to the Lagrangian density of the matter
inside the universe.

In order to describe the $f(T)$ theory we start with  the
following field equations in a FRW background filled with the
pressureless  matter (\cite{ft1})

\begin{equation}
H^2+\frac{k}{a^2}=\frac{1}{3}\,(\rho_m+\rho_T),\label{H}
\end{equation}
\begin{equation}
\dot{H}-\frac{k}{a^2}=-\frac{1}{2}\,(\rho_m+\rho_T+p_T),\label{dH}
\end{equation}
where
\begin{equation}
\rho_T=\frac{1}{2}\,(2Tf'-f-T),\label{rhoT}
\end{equation}
and
\begin{equation}
p_T=-\frac{1}{2}\,[-8\dot{H}T\,f''+(2T-4\dot{H})f'-f+4\dot{H}-T].
\end{equation}

Here,  $H=\dot{a}/a$ is the Hubble factor, $a$ is a scale factor,
$k$ is the curvature parameter, with values -1,0,+1, respectively.
Also, $\rho_m$ is the energy density of the matter,  $\rho_T$ and
$p_T$ are the torsion contributions to the energy density and
pressure. Dots here mean derivatives with respect to the
cosmological times denote the derivative times and  the primes
denotes derivative with respect to the torsion scalar $T$. For
convenience we will use units in which $8\pi G=1$.

Also,the torsion scalar for non-flat background  is defined in
Ref.(\cite{FF}), as
\begin{equation}
T=-6\left(H^2+\frac{k}{a^2}\right).\label{T}
\end{equation}
Considering Eqs.(\ref{H}), (\ref{rhoT}) and (\ref{T}),  the energy
density of the matter can be written as
\begin{equation}
\rho_m=\frac{1}{2}\,[f-2Tf'].\label{rhom}
\end{equation}

On the other hand, following Refs.(\cite{cc,cc2}), we shall assume
that both components, the torsion scalar and the cold dark matter
do not conserve separately but that they interact through a term
$Q$ (to be specified later) according to

\begin{equation}
\dot{\rho_m}+3H\rho_m=\,Q,\label{drm}
\end{equation}
and
\begin{equation}
\dot{\rho_T}+3H(\rho_T+p_T)=\,-Q,\label{rt}
\end{equation}
such that the total energy $\rho=\rho_T+\rho_m$ is conserved i.e.,
$\dot{\rho}+3H(\rho+p)=0$. In what follows we shall consider
$Q>0$. We also assume that the torsion contributions component
obeys an equation of state (EoS) parameter $w_T=p_T/\rho_T$ and
then the Eq.(\ref{rt}) can be written as

\begin{equation}
\dot{\rho_T}+3H\rho_T\left(1+w_T+\frac{Q}{3H\rho_T}\right)=0.\label{rt2}
\end{equation}


Taking time derivative of Eq.(\ref{rhoT}), we get
\begin{equation}
\dot{\rho_T}=\frac{\dot{T}}{2}\,\,[f'+2Tf''-1].\label{ddot}
\end{equation}

Following Ref. (\cite{Ka1}), we combining Eq.(\ref{rt2}) and
(\ref{ddot}) and the EoS parameter yields
\begin{equation}
w_T=-\left[1+\frac{Q}{3H\rho_T}+\frac{\dot{T}}{3H}\,\frac{(2Tf''+f'-1)}{(2Tf'-f-T)}\right],\label{wt2}
\end{equation}
here, we note that the Eq.(\ref{wt2}) represents an effective EoS
parameter.

On the other hand, from Eqs.(\ref{dH}) and (\ref{T}), we get
\begin{equation}
\dot{T}=\frac{12H}{(f'+2Tf'')}\,\left[\frac{(f-2Tf')}{4}+\frac{k}{a^2}\,(f'+2Tf''-1)\right]
.\label{dotT}
\end{equation}
In particular, for $k=0$, $\dot{T}=\frac{3H(f-2Tf')}{(f'+2Tf'')}$,
see Ref. (\cite{KAR}). Substituting Eq.(\ref{dotT}) in
Eq.(\ref{wt2}) the effective EoS parameter results in
$$
w_T=-\Bigl[1+\frac{Q}{3H\rho_T}+\frac{4}{(f'+2Tf'')}\,
\frac{(2Tf''+f'-1)}{(2Tf'-f-T)}\,
$$
\begin{equation}
\left(\frac{(f-2Tf')}{4}+\frac{k}{a^2}\,(f'+2Tf''-1)\right)\Bigr].\label{wtt}
\end{equation}

Note that in a non-interacting limit, i.e., $Q=0$ and $k=0$,  the
effective EoS parameter given by Eq.(\ref{wtt}), reduces to the
standard $f(T)$ gravity, in which
$w_T=-1+4\dot{H}(2Tf''+f'-1)/(2Tf'-f-T)$, see Ref. (\cite{KAR}).



On the other hand, the deceleration parameter $q$ is given by $
q=-\left[1+\frac{\dot{H}}{H^2}\right]$,  and using Eqs.(\ref{H})
and (\ref{dH}) one can obtain

\begin{equation}
q=\frac{1}{2}\,\left[1+\frac{k}{a^2\,H^2}+\frac{\rho_T\,w_T}{H^2}\right].\label{qq}
\end{equation}
Combining Eqs.(\ref{wtt}) and (\ref{qq}) the deceleration
parameter $q$ can be written as
$$
q=\frac{1}{2}-\frac{k}{2a^2}\,\left[\frac{T}{6}+\frac{k}{a^2}\right]^{-1}
+\left[\frac{T}{6}+\frac{k}{a^2}\right]^{-1}\;\times
$$
$$
\Bigl[\frac{(2Tf'-f-T)}{4}+\frac{Q}{6H}+\,
\frac{(2Tf''+f'-1)}{(f'+2Tf'')}\,\times
$$
\begin{equation}
\left(\frac{(f-2Tf')}{4}+\frac{k}{a^2}\,(f'+2Tf''-1)\right)\Bigr].
\label{q}
\end{equation}
We noted that for the particular case in which non-interacting
limit $Q=0$, in the Einstein TG in which $f(T)=T$, and $k=0$, the
above equation gives $q=1/2$, corresponding to the matter
dominated epoch.

\section{GSL interacting - $f(T)$}\label{g}
Having exhibited  the cosmological scenario of a universe
controlled by $f(T)$ gravity, we proceed to an investigation of
its thermodynamic properties, and in particular the validity of
the GSL in a non-flat FRW model occupied with pressureless Dark
Matter (DM), i.e., $p_m=0$. Following Refs.(\cite{S1,S2}) the GSL
the entropy of the horizon plus the entropy of the matter  within
the horizon cannot decrease in time. We suppose that the boundary
of the universe to be enclosed by the dynamical apparent horizon
in FRW universe; accordingly, its radius $R_A$ is given by Ref.
(\cite{RA}), as
\begin{equation}
R_A=\frac{1}{\sqrt{H^2+\frac{k}{a^2}}}\,.\label{RA}
\end{equation}
In particular, for the case $k=0$ the radius of the apparent
horizon coincide with the Hubble horizon, in which $R_A=1/H$, see
Ref. (\cite{Is}).

Following Ref. (\cite{hor}), the Hawking temperature on the
apparent horizon the radius $R_A$ is given by
\begin{equation}
T_A=\frac{1}{2\pi\,R_A}\,\left(1-\frac{\dot{R_A}}{2HR_A}\right).\label{TAA}
\end{equation}
Note that the ratio $\dot{R_A}/2HR_A<1$, ensures that the Hawking
temperature $T_A>0$.

Considering the Gibb's equation, the entropy of the universe
considering DM  inside the apparent horizon $S_m$, is given by
(\cite{Pv})
\begin{equation}
T_A\,dS_m=dE_m+p_m\,dV=dE_m,\label{TA}
\end{equation}
where the volume of the pressureless matter is $V=4\pi\,R_A^3/3$
and
\begin{equation}
E_m=V\,\rho_m=\frac{4\pi\,R_A^3}{3}\,\rho_m.\label{Em}
\end{equation}
From Eqs.(\ref{drm}), (\ref{TA}) and (\ref{Em}), the Gibb's
equation due to the matter, can be written as
\begin{equation}
T_A\,\dot{S}_m=4\pi\,R_A^2\,\rho_m\left(\dot{R}_A+H\,R_A\left[\frac{Q}{3H\rho_m}-1
\right]\right),\label{Sm}
\end{equation}
where $\dot{S}_m$ represent the time derivative of the entropy due
to the matter source inside the horizon. Also, we noted that in
the non-interacting limit the above equation reduces to the Gibb's
equation
$T_A\,\dot{S}_m=4\pi\,R_A^2\,\rho_m\left(\dot{R}_A-H\,R_A\right)$.

Combing Eqs.(\ref{rhom}) and (\ref{Sm}) we get
$$
T_A\,\dot{S}_m=2\pi\,R_A^2\,[f-2Tf']\times
$$
\begin{equation}
\left(\dot{R}_A+
H\,R_A\left[\frac{2\,Q}{3H[f-2Tf']}-1\right]\right).\label{Sm2}
\end{equation}
Here, we noted that in the limit $Q=0$, Eq.(\ref{Sm2}) reduces to
expression obtained in Ref. (\cite{KAR}), in which
$T_A\dot{S}_m=2\pi\,R_A^2(f-2Tf')\,(\dot{R}_A-HR_A)$. Also, we
observed that the evolution of the matter entropy $T_A\dot{S}_m$,
increase with the introduction of the interaction term $Q$.

On the other hand, the contribution of the apparent horizon
entropy $S_A$, in the framework of $f(T)$ gravity when $f''$ is
small, according to Ref. (\cite{Miao}), is obtained as
\begin{equation}
S_A=\frac{A\,f'}{4\,G},\label{SAA}
\end{equation}
where $A=4\pi\,R_A^2$ is the area of the horizon.

In this way, the evolution of horizon entropy considering
Eqs.(\ref{TAA}) and (\ref{SAA}) one can get
\begin{equation}
T_A\,\dot{S}_A=4\pi\left(1-\frac{\dot{R}_A}{2HR_A}\right)\;
\left(2\,\dot{R}_A\,f'+R_A\,f''\,\dot{T}\right),\label{SA}
\end{equation}
and now combining Eqs.(\ref{dotT}) and (\ref{SA}) we get
$$
T_A\,\dot{S}_A=4\pi\left(1-\frac{\dot{R}_A}{2HR_A}\right)\;
\Bigl(2\,\dot{R}_A\,f'+6H\,R_A\,f''
$$
\begin{equation}
\,\times\left[\frac{(f-2Tf')/2-2k/a^2}{(f'+2Tf'')}+\frac{2k}{a^2}\right]
\Bigr).\label{SA2}
\end{equation}
Note that Eq.(\ref{SA2}) coincides with the evolution of the
horizon entropy calculated in Ref. (\cite{Ghosh}) and for a flat
FRW agrees with Ref. (\cite{KAR}). Also we noted that the
evolution of the horizon entropy becomes independent of the
interacting term $Q$.

Taking time derivative of $R_A$ given by Eq.(\ref{RA}) and
considering Eqs.(\ref{dH}) and (\ref{T}), we get
$$
\dot{R_A}=-\frac{1}{2}\,\left(\frac{6}{T}\right)^{3/2}\,\left[\frac{T}{6}+\frac{k}{a^2}\right]^
{1/2}\,\times
$$
\begin{equation}
\left[\frac{(f-2Tf')}{2(f'+2Tf'')}+\frac{k}{a^2}\right].\label{dRA}
\end{equation}
In particular, for the flat FRW metric in which $k=0$, the value
of $\dot{R}_A$ reduces to
$\dot{R}_A=-(3/2T)[(f-2Tf')/(f'+2Tf'')]$, see Ref.\cite{KAR}.

Finally, we need to calculate the total entropy $S_{Total}$ due to
different  contributions of the apparent horizon entropy and the
matter entropy, i.e., $S_{Total}=S_A+S_m$. In this form, the
evolution of the total entropy or rather the GSL, adding
Eqs.(\ref{Sm2}) and (\ref{SA2}), can be written as
$$
T_A\dot{S}_{Total}=2\pi R_A^2 \,
\Bigl[\,\left(\frac{2}{R_A^2}-\frac{\dot{R}_A}{HR_A^3}\right)\;\times
$$
$$
\left(2\,\dot{R}_A\,f'+6HR_Af''\left[\frac{(f-2Tf')/2-2k/a^2}{(f'+2Tf'')}+\frac{2k}{a^2}\right]
\right)
$$
\begin{equation}
+[f-2Tf']\left(\dot{R}_A+
H\,R_A\left[\frac{2\,Q}{3H[f-2Tf']}-1\right]\right)\,\Bigr],\label{ST}
\end{equation}
where $R_A$ and $\dot{R_A}$ are given by Eqs.(\ref{RA}) and
(\ref{dRA}) respectively. Here, we noted that the interacting-term
$Q$ modified  the evolution of the total entropy, results in an
increases on the GLS of thermodynamic, by a factor
$4\pi\,R_A^3\,Q/3>0$. Also, we noted that in the Einstein TG,
where $f(T)=T$, considering a flat universe and in the limit
$Q=0$, the GSL from Eq.(\ref{ST}) results in
$T_A\,\dot{S}_{Total}=9\,\pi>0$, which always prevails (recalled,
that $8\pi\,G=1$).

In the following, we will analyze  the GSL of thermodynamic for a
flat FRW universe, i.e., $k=0$, one specific interaction term $Q$
and one viable model for $f(T)$ with an exponential dependence on
the torsion.

\section{An example for $Q$ and $f(T)$}\label{e}

Let us consider that the interaction term $Q$ is related to the
total energy density of matter and takes the form
(\cite{Rh1,Sdc1,intQ})
\begin{equation}
Q=3\,c^2\,H\,\rho_m,\label{QQ}
\end{equation}
where $c^2$ is a small positive definite constant and the factor 3
was considered for mathematical convenience.

Following Ref. (\cite{9}), we consider one viable model for $f(T)$
with an exponential dependence on the torsion scalar $T$, given by
\begin{equation}
f(T)=T-\mu_2\,T\,\left(1-e^{\beta\,\frac{T_0}{T}}\right),\label{fT}
\end{equation}
where $\mu_2$ and $\beta$ are constants and $T_0=-6H_0$. Note that
the last term of $f(T)$ is analogous to a $f(R)$ model in which an
exponential dependence on the curvature scalar $R$ is considered
(\cite{Li}). The astronomical data from SNeIa+BAO+CMB gives the
following fit values $\beta=-0.02_{-0.20}^{+0.31}$ and
$\Omega_{m0}=0.272_{-0.034}^{+0.036}$ at the 95$\%$ confidence
level (\cite{9}). Here, the dimensionless matter energy
$\Omega_{m0}=8\pi G\rho_{m0}/3H_0^2$ and the parameter
$\mu_2=\frac{1-\Omega_{m0}}{1-(1-2\beta)e^{\beta}}$. In
particular, in the non interacting limit, the values $\beta=0$
reduces to the $\Lambda$CDM. Also, a cosmographic analysis to
check the viability of model given by Eq.(\ref{fT}), was developed
in Ref.(\cite{KAR}) (see also Ref. (\cite{ca})).

Considering Eqs.(\ref{wtt}) and (\ref{QQ})  the effective EoS
parameter, $w_T$, for a flat FRW universe, yields

$$
w_T=-\Bigl[1+\frac{c^2\,(f-2Tf')}{(2Tf'-f-T)}+\frac{(f-2Tf')}{(f'+2Tf'')}\,\times
$$
\begin{equation}
\frac{(2Tf''+f'-1)}{(2Tf'-f-T)}\Bigr],\label{wt1}
\end{equation}
where $f'=1+\mu_2\left(e^{\beta\,\frac{T_0}{T}}\left[\frac{T-\beta
T_0}{T}\right]-1\right)$ and
$f''=\mu_2\beta^2\,T_0^2\,e^{\beta\,\frac{T_0}{T}}/T^3$,
respectively. Here, we noted that in the limit $c^2\rightarrow 0$,
Eq.(\ref{wt1}) reduces to expression calculated in Ref.
(\cite{KAR}).

The acceleration parameter $q$, from Eq.(\ref{q}) is given by

$$
q=\frac{1}{2}+\frac{3}{2\,T}\Bigl[(2Tf'-f-T)+(f-2Tf')\times
$$
\begin{equation}
\left(c^2+ \frac{(2Tf''+f'-1)}{(f'+2Tf'')}\right)\Bigr].\label{q2}
\end{equation}
As before, we noted that in the non interacting limit, i.e.,
$c^2\rightarrow 0$, the parameter $q$ given by Eq.(\ref{q2})
reduces to $q=2[f'-Tf''-(3f/4T)]/(f'-2Tf'')$ and  coincide with
the obtained in Ref. (\cite{KAR}).

The total entropy  due to different contributions of the apparent
horizon entropy and the matter entropy from Eq.(\ref{ST}), can be
written as
$$
T_A\dot{S}_{Total}=2\pi R_A^2 \,
\Bigl[\,\left(\frac{2-\dot{R}_A}{R_A^2}\right)\;
\Bigl(2\,\dot{R}_A\,f'+3\,f''
$$
\begin{equation}
\left[\frac{(f-2Tf')}{(f'+2Tf'')}\right]
\Bigr)+[f-2Tf']\left(\dot{R}_A+
\left[c^2-1\right]\right)\Bigr],\label{fin}
\end{equation}
where $\dot{R}_A$, from Eq.(\ref{dRA}) is given by
$\dot{R}_A=-(3/2T)\,[(f-2Tf')/(f'+2Tf'')]$ and $R_A^2=-6/T$. We
noted that in the Einstein TG, in which, $f(T)=T$, the GSL from
Eq.(\ref{fin}) results in $T_A\,\dot{S}_{Total}=\pi\,(9+12c^2)>0$,
which always prevails and also we noted that the GLS increases
with the interaction-term (recalled, that $8\pi\,G=1$).

\begin{figure}[th]
\includegraphics[width=3.in,angle=0,clip=true]{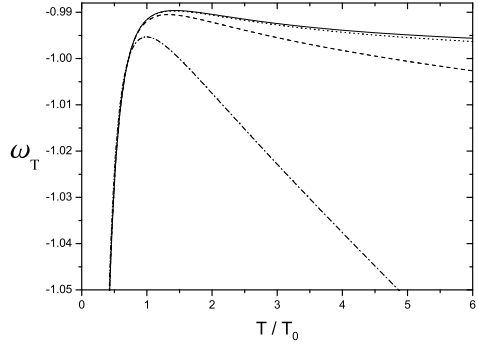}
\caption{Evolution of the effective EoS parameter $w_T$ versus the
dimensionless torsion $T/T_0$, for different values of the
interaction parameter $c^2$. Dot-dashed, dashed, dotted and solid
lines are for the interaction parameter $c^2=10^{-2}$,
$c^2=10^{-3}$, $c^2=10^{-4}$ and  the non-interacting limit
$c^2=0$, respectively. Here, we have used $H_0=74.2$ Km S$^{-1}$,
$\beta=-0.02$ and $\mu_2=-37.5$\label{fig1}}
\end{figure}

In Fig.(\ref{fig1}) we show the evolution from the early times
($T/T_0\rightarrow +\infty$) to the current  epoch ($T/T_0=1$) for
the effective EoS parameter $w_T$ versus the dimensionless torsion
$T/T_0$, for different values of the interaction parameter $c^2$.
Dot-dashed, dashed, dotted and solid lines are for the interaction
parameter $c^2=10^{-2}$, $c^2=10^{-3}$, $c^2=10^{-4}$ and  the
non-interacting limit $c^2=0$, respectively. In order to write
down values that relate the effective EoS versus $T/T_0$, we
considered  Eq.(\ref{wt1}). Also, we have used the best fit values
$\beta=-0.02$ and $\Omega_{m0}=0.272$  (\cite{9}), in which
$\mu_2=\frac{1-\Omega_{m0}}{1-(1-2\beta)e^{\beta}}=-37.5$ and
$H_0=74.2$ Km S$^{-1}$ (\cite{Ri}). An attractive result which is
not present in the non interacting limit model($c^2=0$) in the
past (\cite{KAR,9}), is that for the different  values of the
interacting parameter $c^2$, we have a transition from the
$w_T>-1$ (quintessence) to $w_T<-1$ (phantom). In this form, the
interacting $f(T)$ gravity can cross the phantom divide line, as
could be seen from Fig.(\ref{fig1}). In particular, this
transition occurs for the parameter $c^2=10^{-2}$ at $T/T_0=1.5$
and for  $c^2=10^{-3}$ at $T/T_0=4.74$. We have found that the
value $c^2=10^{-4}$ (dotted line) presents a small displacement in
relation to the non interacting limit that corresponds to $c^2=0$.
At the present epoch, in particular for the values $c^2=10^{-2}$
and $c^2=10^{-3}$, we find that $w_{T_0}=-0.995$ and
$w_{T_0}=-0.992$, respectively. In the future in which
$1<T/T_0<0$, again we have a transition from $w_T>-1$ to $w_T<-1$,
for the different values of $c^2$. This result has been previously
noted in Ref. (\cite{KAR}), for the case $c^2=0$ in which
$T/T_0=0.719$.

\begin{figure}[th]
\includegraphics[width=3.in,angle=0,clip=true]{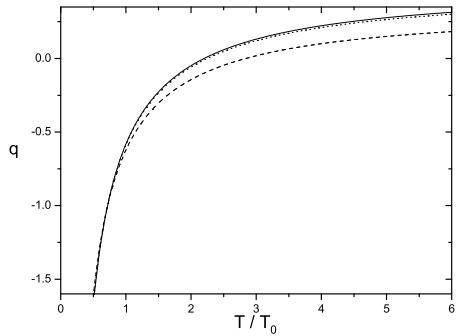}
\caption{The plot shows the evolution of the deceleration
parameter $q$ versus the dimensionless torsion $T/T_0$, for three
different values of the interaction parameter $c^2$. Here, dashed,
dotted and solid lines are for the interaction parameter
$c^2=10^{-1}$, $c^2=10^{-2}$ and  the non-interacting limit
$c^2=0$, respectively. As before, we have used $H_0=74.2$ Km
S$^{-1}$, $\beta=-0.02$ and $\mu_2=-37.5$. \label{fig2}}
\end{figure}

In Fig.(\ref{fig2}) we show the evolution  of the deceleration
parameter $q$  versus the dimensionless torsion $T/T_0$, for three
different values of the interaction parameter $c^2$. Dashed,
dotted and solid lines are for the interaction parameter
$c^2=10^{-1}$, $c^2=10^{-2}$ and  the non-interacting limit
$c^2=0$, respectively. In order to write down values that relate
$q$ versus $T/T_0$, we used Eq.(\ref{q2}). As before, we have used
the values $\beta=-0.02$ and $\Omega_{m0}=0.272$, then
$\mu_2=\frac{1-\Omega_{m0}}{1-(1-2\beta)e^{\beta}}=-37.5$ and
$H_0=74.2$ Km S$^{-1}$. From Fig.(\ref{fig2}) we have a cosmic
transition from $q>0$ (deceleration) to $q<0$ (acceleration) with
is consistent with the observations Ref. (\cite{dali}). In
particular, this transition occurs for the parameter $c^2=10^{-1}$
at $T/T_0=2.84$ and for $c^2=10^{-2}$ at $T/T_0=2.2$,
respectively. At the early times, i.e., $T/T_0\rightarrow\infty$,
for the value $c^2=10^{-1}$ we find that the deceleration
parameter $q_\infty=0.35$ and for $c^2=10^{-2}$ corresponds to
$q_\infty=0.5$ and this value of the deceleration parameter at the
early times, coincides  with the non-interacting limit $c^2=0$. At
the current epoch i.e., $T/T_0=1$, we obtain that for the
interacting parameter $c^2=10^{-1}$ corresponds to $q_0=-0.62$ and
for $c^2=10^{-2}$ corresponds to $q_0=-0.58$ and these values are
in accord with the observational result Ref. (\cite{dali}).

\begin{figure}[th]
\includegraphics[width=3.in,angle=0,clip=true]{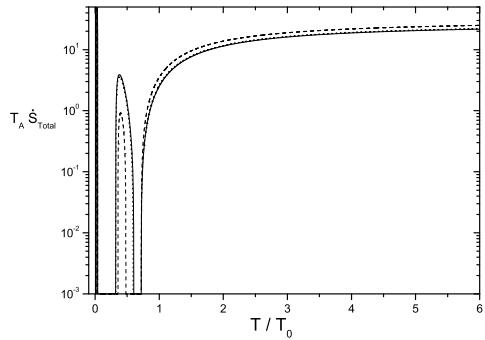}
\caption{ The plot shows the evolution of the GLS
($T_A\dot{S}_{Total}$) versus $T/T_0$, for three different values
of the interaction parameter $c^2$.  As before, dashed, dotted and
solid lines are for the interaction parameter $c^2=10^{-1}$,
$c^2=10^{-2}$ and $c^2=0$, respectively. Here, as before we have
used $H_0=74.2$ Km S$^{-1}$, $\beta=-0.02$ and $\mu_2=-37.5$.
Here, we have considered a logarithmic scale for
$T_A\dot{S}_{Total}$. \label{fig3}}
\end{figure}

In Fig.(\ref{fig3}) we show the evolution  of the GLS   versus the
dimensionless torsion $T/T_0$, for three different values of the
interaction parameter $c^2$. Dashed, dotted and solid lines are
for the interaction parameter $c^2=10^{-1}$, $c^2=10^{-2}$ and
 $c^2=0$, respectively. In order to write
down values that relate $T_A\dot{S}_{Total}$ versus $T/T_0$, we
considered  Eq.(\ref{fin}). As before, we have used $\beta=-0.02$
and $\Omega_{m0}=0.272$, then
$\mu_2=\frac{1-\Omega_{m0}}{1-(1-2\beta)e^{\beta}}=-37.5$ and
$H_0=74.2$ Km S$^{-1}$. From Fig.(\ref{fig3}) we observed that the
GLS is satisfied from the early times i.e.,
$T/T_0\rightarrow\infty$ to the current epoch in which $T/T_0=1$.
Also, we noted that the GLS graphs for the value $c^2=10^{-2}$
present a small displacement with respect to the dimensionless
torsion $T/T_0$, when compared to the results obtained in the
non-interacting limit model, in which $c^2=0$. In particular, at
early times, i.e., $T/T_0\rightarrow\infty$ we obtain that
$T_A\dot{S}_{Total}\simeq 25.01 $ for the value $c^2=10^{-1}$,
$T_A\dot{S}_{Total}\simeq22.04$ that corresponds to $c^2=10^{-2}$
and  $T_A\dot{S}_{Total}\simeq21.71$ that corresponds to $c^2=0$
i.e, the non-interacting limit. At the present time i.e.,
$T/T_0=1$ we get that $T_A\dot{S}_{Total}\simeq 2.61 $ for the
value $c^2=10^{-1}$, $T_A\dot{S}_{Total}\simeq2.61$ that
corresponds to $c^2=10^{-2}$ and  $T_A\dot{S}_{Total}\simeq2.51$
that corresponds to $c^2=0$. Analogously, as occurs in the
non-interacting limit, the GLS is infringed  in the future i.e.,
$1<T/T_0<0$, in which GLS is negative, $T_A\dot{S}_{Total}<0$. In
particular, for the value $c^2=10^{-2}$ the GLS is violated for
the dimensionless torsion $T/T_0$, in the special intervals
(0.005,0.017), (0.038,0.356) and (0.480,0.719).

\section{Conclusions \label{conclu}}

In this paper we have investigated the GLS in the context of
$f(T)$ gravity. We studied the GLS from the boundary of the
universe to be enclosed by the dynamical apparent horizon in
non-flat FRW universe occupied with pressureless DM, together with
the Hawking temperature on the apparent horizon. We found that the
the interacting term $Q$ modified; the torsion contributions
component given by an effective EoS parameter, the deceleration
parameter and the  evolution of the total entropy or rather the
GSL. In particular, we noted that the modified in the evolution of
the total entropy, results in an increases on the GLS of
thermodynamic by a factor $4\pi\,R_A^3\,Q/3>0$.

Our specific model  is described by a viable model $f(T)$ with an
exponential dependence on the torsion $T$ in a flat FRW universe
and have considered for simplicity the case in which the
interaction term $Q$ is related to the total energy density of
matter. For this model, we have also obtained explicit expressions
for the effective EoS parameter, the deceleration parameter and
the evolution of the total entropy. For these particular choices
of the $f(T)$ and $Q$, it is possible to obtain an accelerating
expansion of the universe (here, the effective torsion fluid
representing  the role of dark energy). We also found an
attractive result that which is not present in the non interacting
limit model($c^2 = 0$) in the past, is that for values different
of the interacting parameter $c^2\neq0$, we obtained a transition
from the $w_T > 1$ (quintessence) to $w_T < 1$ (phantom). In this
form, a crossing of phantom divide line is possible  for the
interacting-$f(T)$ model in the past. Also, we found  a cosmic
transition in the past from $q
> 0$ (deceleration) to $q < 0$ (acceleration) with is consistent
with the observations. Here, we noted that for values of
$c^2<10^{-2}$ the value of the deceleration parameter at the early
times, coincides with the non-interacting limit $c^2 = 0$ and
becomes $q_\infty=0.5$. For our specific model, we analyzed the
validity of the GLS and we found an increases on the total entropy
by a factor $2\pi\,R_A^2\,(f-2Tf')\,c^2$ product of the
interaction-term. Also, we obtained that $T_A\dot{S}_{Total}>0$
from the early times to the current epoch. However, in the future
($1<T/T_0<0$) the total entropy becomes $T_A\dot{S}_{Total}<0$ and
then the GLS is violated, for three special intervals of the
dimensionless torsion $T/T_0$, independently of the
interaction-term.

Finally, we have shown that the GLS of thermodynamics for the
interacting in $f(T)$ gravity is less restricted than analogous
$Q=0$ due to the introduction of a new parameter, proportionates
for the interaction term $Q$. In our specific model the
incorporation of this parameter gives us a freedom that allows us
to modify the standard $f(T)$ gravity by simply modifying the
corresponding value of the parameter $c^2$ associated to the
interaction term $Q$. Also, we have not addressed other
interacting-$f(T)$ models. We hope to return to this point in near
future.

\begin{acknowledgments}
This work  was supported by COMISION NACIONAL DE CIENCIAS Y
TECNOLOGIA through FONDECYT Grant  N$^0$ 1130628 and by DI-PUCV
Grant 123724.
\end{acknowledgments}

\end{document}